\documentclass[11pt]{article}
\usepackage{a4p,cite}
\usepackage[dvips]{graphicx}
\newcommand{\epem}  {\ensuremath{\mathrm{e^+e^-}}}
\newcommand{\zb}    {\mbox{$\rm Z^0$}}
\newcommand{\qq}    {\ensuremath{\mathrm{q\bar{q}}}}
\newcommand{\bb}    {\ensuremath{\mathrm{b\bar{b}}}}
\newcommand{\cc}    {\ensuremath{\mathrm{c\bar{c}}}}
\newcommand{\light} {\ensuremath{\mathrm{u\bar{u},d\bar{d},s\bar{s}}}}
\newcommand{\n}     {\ensuremath{\bar{n}}}
\newcommand{\nb}    {\ensuremath{\bar{n}_{\rm b\bar{b}}}}
\newcommand{\nc}    {\ensuremath{\bar{n}_{\rm c\bar{c}}}}
\newcommand{\nl}    {\ensuremath{\bar{n}_{\rm l\bar{l}}}}
\newcommand{\dbl}   {\ensuremath{\delta_{\rm bl}}}

\newcommand{\naive} {na\"{\i}ve} 
\newcommand{\NQdec}{N_{\mathrm{Q}}^{\mathrm{decay}}}
\newcommand{\xQ}{x_{\mathrm{Q}}} 
\newcommand{\xQb}{x_{\overline{\mathrm{Q}}}}  
\begin{document}
\begin{titlepage}
\noindent
\begin{center}  {\large EUROPEAN ORGANIZATION FOR NUCLEAR RESEARCH }
\end{center}
\bigskip\bigskip\bigskip
\begin{tabbing}
\` CERN-EP/2002-079\\
\` October 18, 2002
\end{tabbing}
\vspace{1.0cm}

\begin{center}
    \LARGE\bf\boldmath
Charged particle multiplicities in heavy and light quark initiated events    
above the $\rm Z^0$ peak
\end{center}
\vspace{1.0cm}
\begin{center}
{\Large\bf
The OPAL Collaboration
}
\end{center}
\vspace{1.5cm}
\begin{abstract}

We have measured the mean charged particle multiplicities separately
for $\bb$, $\cc$ and light quark ($\light$) initiated events produced
in {\epem} annihilations at LEP.
The data were recorded with the OPAL detector at eleven different
energies above the \zb\ peak,   
corresponding to the full statistics collected at LEP1.5 and LEP2.

The difference in mean charged particle multiplicities for $\bb$ and light quark 
events, $\dbl$, measured over this energy range is 
consistent with an energy independent behaviour, as predicted by QCD, but 
is inconsistent with the prediction of a more phenomenological approach 
which assumes that the multiplicity accompanying the decay of a heavy quark 
is independent of the quark mass itself. 
Our results, which can be combined into the single measurement 
$\dbl = 3.44 \pm 0.40$~(stat)~$\pm 0.89$~(syst) at a luminosity weighted average 
centre-of-mass energy of 195~GeV, are also consistent with an energy independent 
behaviour as
extrapolated from lower energy data.
 
\end{abstract}
\vspace{1.0cm}
\begin{center} 
{\Large (Submitted to Physics Letters B) }
\end{center}

\end{titlepage}
\begin{center}{\Large        The OPAL Collaboration
}\end{center}\bigskip
\begin{center}{
G.\thinspace Abbiendi$^{  2}$,
C.\thinspace Ainsley$^{  5}$,
P.F.\thinspace {\AA}kesson$^{  3}$,
G.\thinspace Alexander$^{ 22}$,
J.\thinspace Allison$^{ 16}$,
P.\thinspace Amaral$^{  9}$, 
G.\thinspace Anagnostou$^{  1}$,
K.J.\thinspace Anderson$^{  9}$,
S.\thinspace Arcelli$^{  2}$,
S.\thinspace Asai$^{ 23}$
% ...
}\end{center}\bigskip
\bigskip
$^{  1}$School of Physics and Astronomy, University of Birmingham,
Birmingham B15 2TT, UK
\newline
$^{  2}$Dipartimento di Fisica dell' Universit\`a di Bologna and INFN,
I-40126 Bologna, Italy
\newline
$^{  3}$Physikalisches Institut, Universit\"at Bonn,
D-53115 Bonn, Germany
\newline
% ...
$^{  5}$Cavendish Laboratory, Cambridge CB3 0HE, UK
\newline
% ...
$^{  9}$Enrico Fermi Institute and Department of Physics,
University of Chicago, Chicago IL 60637, USA
\newline
% ...
$^{ 16}$Department of Physics, Schuster Laboratory, The University,
Manchester M13 9PL, UK
\newline
% ...
$^{ 22}$Department of Physics and Astronomy, Tel Aviv University,
Tel Aviv 69978, Israel
\newline
$^{ 23}$International Centre for Elementary Particle Physics and
Department of Physics, University of Tokyo, Tokyo 113-0033, and
Kobe University, Kobe 657-8501, Japan
\newline
% ...
%
\section{Introduction}
The study of heavy quark pair production in \epem collisions with centre-of-mass 
energies greatly exceeding the heavy 
quark masses provides important tests of perturbative QCD. 
In particular mass effects 
are expected to induce substantial differences in the accompanying soft gluon  
radiation in heavy or light quark initiated events.
A prediction of QCD concerns
the difference in charged particle multiplicity, $\dbl$, between $\bb$ events and 
light quark $\rm l\bar{l}$~($\equiv \light$) events, which is expected 
to be almost energy independent~\cite{bib-schumm,bib-petrov,bib-dedeus}.
For a recent review see for example~\cite{bib-khoze-ochs}.
This prediction is in striking contrast with that 
from a more phenomenological approach, the so-called \naive\  
model~\cite{bib-rowson,bib-naive-model}, which assumes that the hadron 
multiplicity accompanying the heavy hadrons in $\bb$ events is the same as the 
multiplicity in light quark events at the energy left to the system once 
the heavy quarks have fragmented.  
This \naive\ model predicts that $\dbl$ decreases with 
increasing centre-of-mass energy.

Experimental results published at 
$\sqrt{s} = 91$~GeV~\cite{bib-mark2,bib-opal-lep1,bib-sld,bib-delphi-lep1}
and at lower energies~\cite{bib-rowson,bib-low-e-dbl}, were not conclusive. 
Since the difference between the two theoretical predictions increases with 
increasing energy, a more powerful discrimination can be attempted at LEP2.
Unfortunately the numbers of events collected at these higher
energies are much smaller than at LEP1, and the results will therefore be 
much less precise. 

In this analysis we have measured the mean charged particle 
multiplicity separately for b, c and uds initiated events
observed with the OPAL detector at all eleven LEP energies above the 
\zb\ resonance, ranging from $\sqrt{s} = 130$~GeV to $\sqrt{s} = 206$~GeV,
and derived at each energy the difference $\dbl$.
This independent set of measurements covers fairly uniformly   
an energy interval of almost 80~GeV, and 
provides a clear discrimination between the two theoretical predictions.   

The DELPHI Collaboration has published its first 
results~\cite{bib-delphi-lep2} at three LEP2 energies, $\sqrt{s} = 183$, 189 
and 200 GeV.
Their measured values of $\dbl$ were found to be consistent with an 
energy independent extrapolation from lower energy data, and inconsistent with 
the prediction of the \naive\ model by more than three standard deviations. 

\section{Data sample and event simulation}
The OPAL detector has been described in detail elsewhere~\cite{bib-OPALdet}. 
The analysis presented here relies mainly on the reconstruction of charged 
particles in the central detector, which consisted of a silicon microvertex 
detector, a precision vertex drift chamber, a large volume jet chamber 
and drift chambers measuring the coordinate along the beam axis as they
leave the jet chamber\footnote{The coordinate system of OPAL has the $z$ 
axis along the electron
beam direction, the $y$ axis pointing upwards and $x$ towards the centre of the
LEP ring. The polar angle $\theta$ is measured with respect to the $z$ axis.}. 
A solenoid providing a field of 0.435\,T along the beam
axis surrounded all tracking detectors. 

The analysis is based on data recorded with the OPAL detector between 1995 and
2000 at eleven different centre-of-mass energies, namely $\sqrt{s} = 130$, 136, 
161, 172, 183, 189, 192, 196, 200, 202 and 206~GeV.
The data recorded in the year 2000 were mostly taken between 205 GeV and 207 GeV,
with a weighted mean value of 206.1 GeV, and were analysed as a single energy.
The data at 130 and 136~GeV were recorded in two different years, 1995 and 1997.
We have checked, at both energies, that the two data sets give completely consistent 
results within the expected statistical uncertainties and therefore we have combined them.
The integrated luminosity collected at the different centre-of-mass energies is
detailed in Table~\ref{table1}.

High statistics samples of Monte Carlo events were generated at each
energy to simulate the relevant physics process and 
the potential background. 
All generated events were passed through a detailed simulation of the
OPAL detector~\cite{bib-gopal} and processed using the same reconstruction
and selection algorithms as the real data.
To simulate signal events of the type 
$\epem \rightarrow \zb / \gamma^* \rightarrow \qq$,
we used the PYTHIA 5.7 parton shower model with fragmentation
provided by the JETSET 7.4 routines~\cite{bib-pythia57} up to
189 GeV, and the PYTHIA 6.1 Monte Carlo program~\cite{bib-pythia61}
interfaced with the KK2f program~\cite{bib-kk2f} at higher energies to obtain
a more accurate description of initial state radiation.
Both models have been tuned to describe the OPAL data at the \zb\ peak
energy~\cite{bib-opaltune}.
As an alternative fragmentation model we used events generated with
HERWIG 6.2~\cite{bib-herwig-62} interfaced to KK2f, also
tuned to the OPAL data\footnote{The main changes from the default tune are that
meson states that do not belong to the $L=0,1$ supermultiplets
are removed, and that the parameters CLSMR(1), PSPLT(2) and DECWT 
have been changed from their default values of 0.0, 1.0, and 1.0 to 
0.4, 0.33, and 0.7. A detailed description of the tune can be 
obtained from the HERWIG web interface.}.
We have observed that version 6.2 of HERWIG provides a substantial improvement 
with respect to version 5.9 in the description of the heavy quark sector, in particular 
the charged particle multiplicities. 
Both the PYTHIA and HERWIG models now provide an adequate description of multihadronic 
final states up to the highest LEP energies. 
Two-photon processes were simulated using PYTHIA, HERWIG and 
PHOJET~\cite{bib-phojet}, and $\tau$ pair production using KORALZ~\cite{bib-koralz}.
The 4-fermion background was studied using high statistics samples of 
Monte Carlo events generated with the grc4f 2.1 model~\cite{bib-grace4f}, interfaced to 
JETSET 7.4 using the same parameter set for the parton shower, fragmentation and decays 
as mentioned above.
 
\section{Event and track selection}

The selection of non-radiative {\qq} events was the same as in previous 
OPAL analyses~\cite{bib-QCD-standard,bib-wqcd}.
In a first step, hadronic events were identified using criteria  
described in~\cite{OPALPR035}, optimised for running at energies above 
the \zb\ peak.
The efficiency for selecting non-radiative hadronic events is greater than 
98\%, where simulated non-radiative events are defined as those with
an invariant mass at the generator level, not considering photons radiated from
the initial state, within 1 GeV of the nominal centre-of-mass energy.

In addition, it was required that charged tracks had transverse momentum 
$p_{T} > 150$~MeV/$c$ with respect to the beam axis, a minimum number of 40 hits in 
the jet chamber, a maximum allowed distance of the point of closest approach to the 
collision point in the $r-\phi$ plane, $d_0$, of 2 cm and that this point should lie 
within 25 cm of the origin in the $z$ direction.

To ensure a good containment in the detector and to reject background 
events of the type $\gamma\gamma\rightarrow \qq$ and $\epem\rightarrow\tau^+\tau^-$, 
we required that the polar angle of the thrust axis $\theta_T$, computed using charged 
tracks which passed the above mentioned criteria, satisfied the condition 
$|\cos\theta_T|<0.9$ and that there were at least seven accepted tracks. 
The residual background from these two processes was estimated to be less 
than 0.3\% and consequently neglected.
To reject events with large initial-state radiation we determined the effective 
centre-of-mass energy of the observed hadronic system, $\sqrt{s'}$, as described 
in~\cite{OPALPR211}. 
Events were rejected if $\sqrt{s'}<\sqrt{s}- 10$~GeV, where $\sqrt{s}$ is the
nominal centre-of-mass energy.  

Above 160 GeV the 4-fermion background becomes significant and is 
dominated by hadronic decays of W pairs.
This background was reduced to a manageble level by cutting on the QCD event
weight variable, $W_{\rm QCD}$, as in~\cite{bib-QCD-standard}.
This variable tests the compatibility of the events with QCD-like processes and details 
of its definition and performance for 4-fermion background rejection can be found 
in~\cite{bib-wqcd}.
We accepted events with $W_{\rm QCD} \geq - 0.5$.
After all cuts we found an overall efficiency for non-radiative {\qq}
events of about 78\%.
The residual fraction of events with a true effective centre-of-mass energy below 
$\sqrt{s} - 10$~GeV is about 5\%.
The estimated residual 4-fermion background varies from 2.4\%, 
at 161 GeV, to 11.6\% at 206 GeV and was
subtracted directly from the observed distributions, as described in the next section.  
The number of events selected at each energy after all cuts is presented in 
Table~\ref{table1}. 
\section{Experimental method and results}
\begin{figure}[tbp]
\centering
\vspace*{-4mm}
\includegraphics[width=1.0\textwidth]{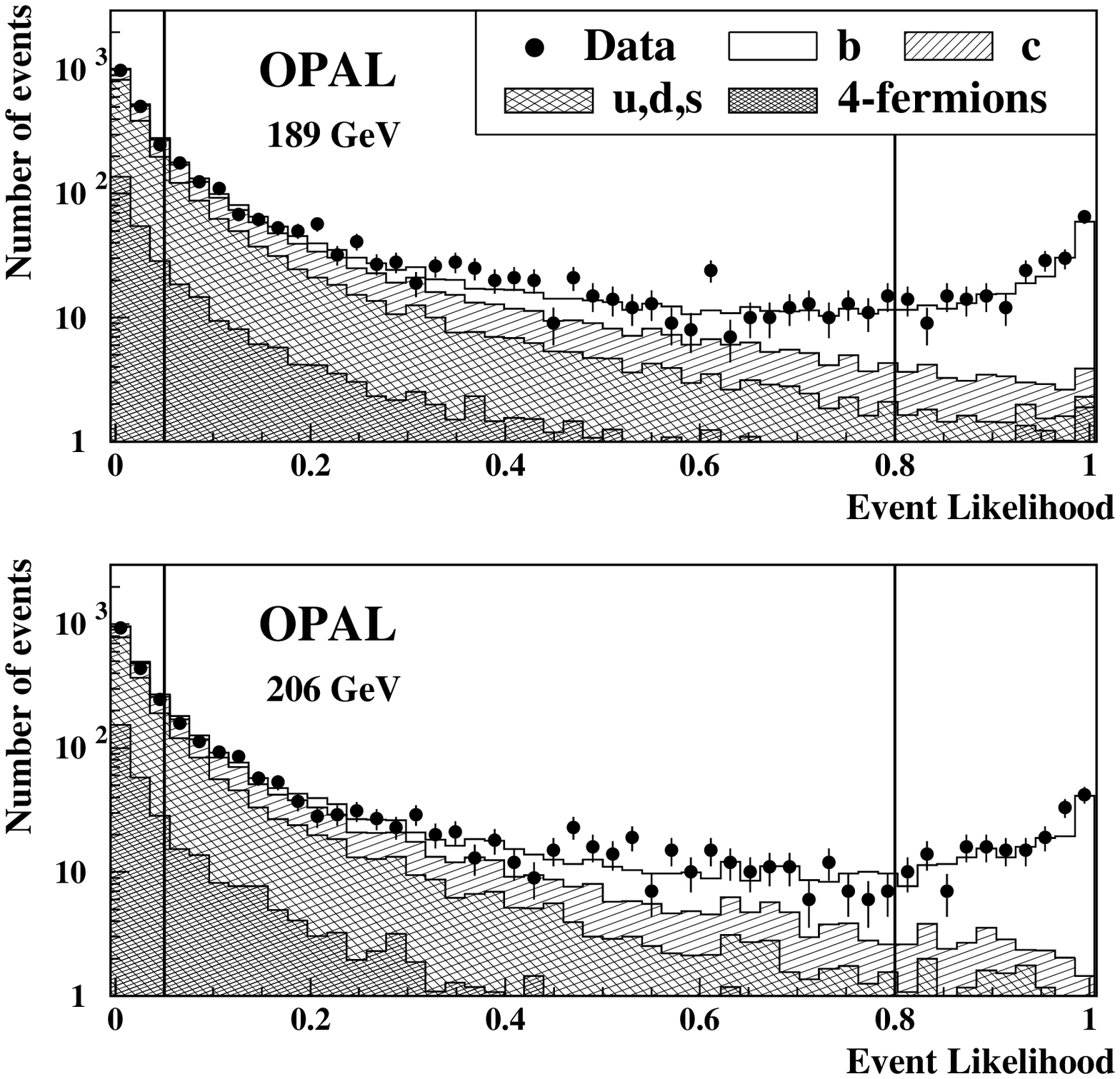}
\caption{\label{likelihood}
The event b-tagging likelihood ${\cal L}$ for all selected events at $\sqrt{s} = 189$ 
and $\sqrt{s} = 206$~GeV.
The histograms show the expectation from Monte Carlo simulation.
The vertical lines define the three independent samples used in this analysis.}
\end{figure}

As in previous OPAL studies at LEP1 energies~\cite{bib-opal-lep1} and 
as proposed for LEP2 energies in~\cite{bib-cern-9601}, we used a method based on 
the simultaneous analysis of event samples with different quark flavour 
compositions to extract charged particle multiplicities separately for each flavor.
At each centre-of-mass energy we selected three independent samples, one highly enriched 
in $\light$ events (Sample~1), one slightly enriched in $\cc$ events with 
respect to an inclusive sample (Sample 2) and one highly enriched in $\bb$ events (Sample 3).
The selection of these three samples was made using a well tested  b-tagging 
technique~\cite{bib-lb-tagging} which uses information from b hadron lifetime, transverse
momentum of leptons with respect to the jet axis, and kinematic observables.
This b-tag uses artificial neural networks (ANNs) to combine optimally
lifetime-sensitive tagging variables and also to combine kinematic variables in the 
jet-kinematics part of the tagging.
The outputs of the lifetime ANN, the kinematic ANN and the lepton tag are finally combined, 
by using an unbinned likelihood method, into a single-valued variable ${\cal L}$, which 
reflects the likelihood that a multihadronic event originates from a $\bb$ pair.
The distribution of the event likelihood ${\cal L}$ at the two energies
with highest statistics is shown in Figure~\ref{likelihood}.
In the same figure the contributions from the different quark flavours 
and the residual 4-fermion background, as predicted from fully simulated
events, are also shown.

\begin{figure}[tbp]
\centering
\vspace*{-4mm}
\includegraphics[width=1.0\textwidth]{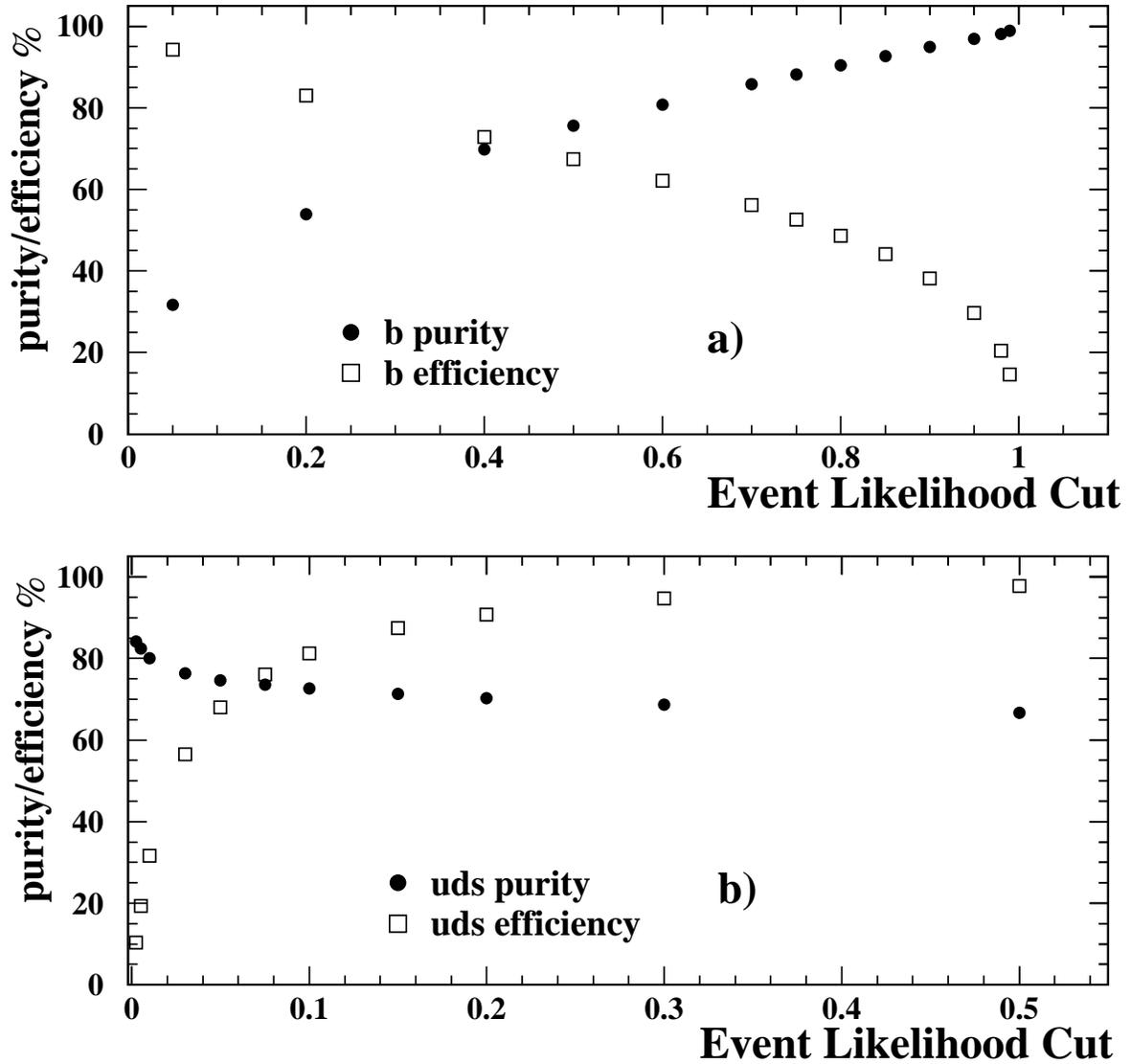}
\caption{\label{flav-vs-bevent}
a) b quark and b) uds quark purity and efficiency (in \%) as a function of 
the event likelihood cut values at  $\sqrt{s} = 189$~GeV.}
\end{figure}

We show in Figure~\ref{flav-vs-bevent} the $\bb$ and $\light$ event purities and 
efficiencies as a function of the cut on the ${\cal L}$ tagging variable, obtained 
from Monte Carlo events generated with PYTHIA at 189 GeV.
The purity for b quarks at a given value X of ${\cal L}$ is defined as 
the fraction of genuine $\bb$ events with ${\cal L}$~$\geq$~X with respect 
to all events tagged with ${\cal L}$~$\geq$~X.
Similarly, for uds quarks the purity at a given value Y of ${\cal L}$ 
is defined as the fraction of genuine $\light$ events 
tagged with ${\cal L}$~$\leq$~Y with respect to all events tagged with 
${\cal L}$~$\leq$~Y. 
The efficiencies are defined as the fractions of b or uds events
tagged with that particular cut value of ${\cal L}$ with respect to the total
number of produced b or uds events.

We used ${\cal L}$ $\geq$ 0.80 to select samples enriched in $\bb$ events 
(Sample 3), and ${\cal L}$  $\leq$ 0.05 to select samples enriched in $\light$ events (Sample 1).
The remaining events, namely those with 0.05~$<$~${\cal L}$~$<$~0.80, comprise
Sample 2.
The two vertical lines shown in Figure~\ref{likelihood} show the cuts used to define the 
three samples. 
We could not select a sample highly enriched in $\cc$ events since the ${\cal L}$ variable is 
not sufficiently sensitive to the c quark fragmentation properties, and the limited statistics 
available at each energy does not allow direct c quark tagging using exclusive reconstruction
of charm mesons, as was done at the \zb\ peak~\cite{bib-opal-lep1}.
The low c quark purity of this sample translates into large uncertainties in the 
measurements of \nc.
Sample 2 contains, however, a slightly higher fraction of $\cc$ events compared to an inclusive 
sample, (30-33\% against 22-26\%, depending on the exact centre-of-mass energy) and has the 
advantage of being completely independent of the other two samples. 

At a given energy, and after the subtraction of the residual 4-fermion background, the mean 
charged particle multiplicity measured in each sample, 
$\n_i^{\rm corr}$ ($i=1,3$), corrected for detector effects, event selection cuts, residual 
contamination of radiative events and biases introduced by the tagging 
procedure, is a linear combination of the unknowns {\nb}, {\nc} and {\nl} (l = u,d,s), the 
true mean multiplicities of the corresponding {\qq} events.
One can extract {\nb}, {\nc} and {\nl} by solving the system of equations
\begin{eqnarray}
\n_1^{\rm corr} & = & f^{\rm b}_1\nb + f^{\rm c}_1\nc + f^{\rm l}_1\nl \nonumber \\
\n_2^{\rm corr} & = & f^{\rm b}_2\nb + f^{\rm c}_2\nc + f^{\rm l}_2\nl \\
\n_3^{\rm corr} & = & f^{\rm b}_3\nb + f^{\rm c}_3\nc + f^{\rm l}_3\nl \nonumber
\end{eqnarray}
where $f^b_i$, $f^c_i$ and $f^l_i$ ($i$ = 1,2,3) are the fractions of $\bb$, 
$\cc$ and $\light$ events in the $i^{th}$ sample, evaluated from fully 
simulated $\epem \rightarrow \zb / \gamma^* \rightarrow \qq$ events.
The experimentally corrected mean charged particle multiplicity is defined 
\cite{bib-cern-8908} as the total number of all promptly produced 
stable charged particles and those produced in the decay of particles with 
lifetimes shorter than $3 \cdot 10^{-10}$ sec. 

For each sample, the multiplicity distribution of the residual 4-fermion background 
after all selection cuts, normalised to the data integrated luminosity, was estimated 
from Monte Carlo events and directly subtracted from the experimentally measured distribution.
We show in Table~\ref{table1} the number of events left after all cuts 
and background subtraction, together with the corresponding 4-fermion background which 
was subtracted.
The fractions of $\bb$, $\cc$ and $\light$ events as 
predicted by PYTHIA are also shown.
The fractions predicted by HERWIG typically differ by around $1\%$.    

To calculate the values of $\n_i^{\rm corr}$, the mean values $\bar{n}_i^{\rm obs}$ of 
the measured distributions were corrected for detector effects, event selection cuts, 
biases introduced by the b-tagging technique and by the residual contamination of radiative 
events by applying a multiplicative correction factor, $C_i$,
\begin{eqnarray}
\n_i^{\rm corr} & = & C_i \cdot \bar{n}_i^{\rm obs}  
\; \; \; \; \; ; \; \; \; \; \; 
C_i  =  \bar{n}_i^{\rm MC-had}/\bar{n}_i^{\rm MC-obs} 
\end{eqnarray}
where $\bar{n}_i^{\rm obs}$ is the observed uncorrected mean charged particle 
multiplicity measured for the $i^{th}$ sample in the data and 
$\bar{n}_i^{\rm MC-obs}$ is the same quantity obtained from a high statistics 
sample of fully simulated events.
$\bar{n}_i^{\rm MC-had}$ is the mean charged particle multiplicity 
obtained from a Monte Carlo sample with the quark flavour 
fractions as expected in sample $i$, without detector simulation 
and without simulation of the initial state radiation process.
We have checked that the accuracy and the precision on the determination of the mean 
values obtained by using the correction method based on a simple multiplicative factor are 
completely equivalent to those obtained by a full matrix unfolding~\cite{bib-multpaper}.

\begin{figure}[tbp]
\centering
\vspace*{-4mm}
\includegraphics[width=1.0\textwidth]{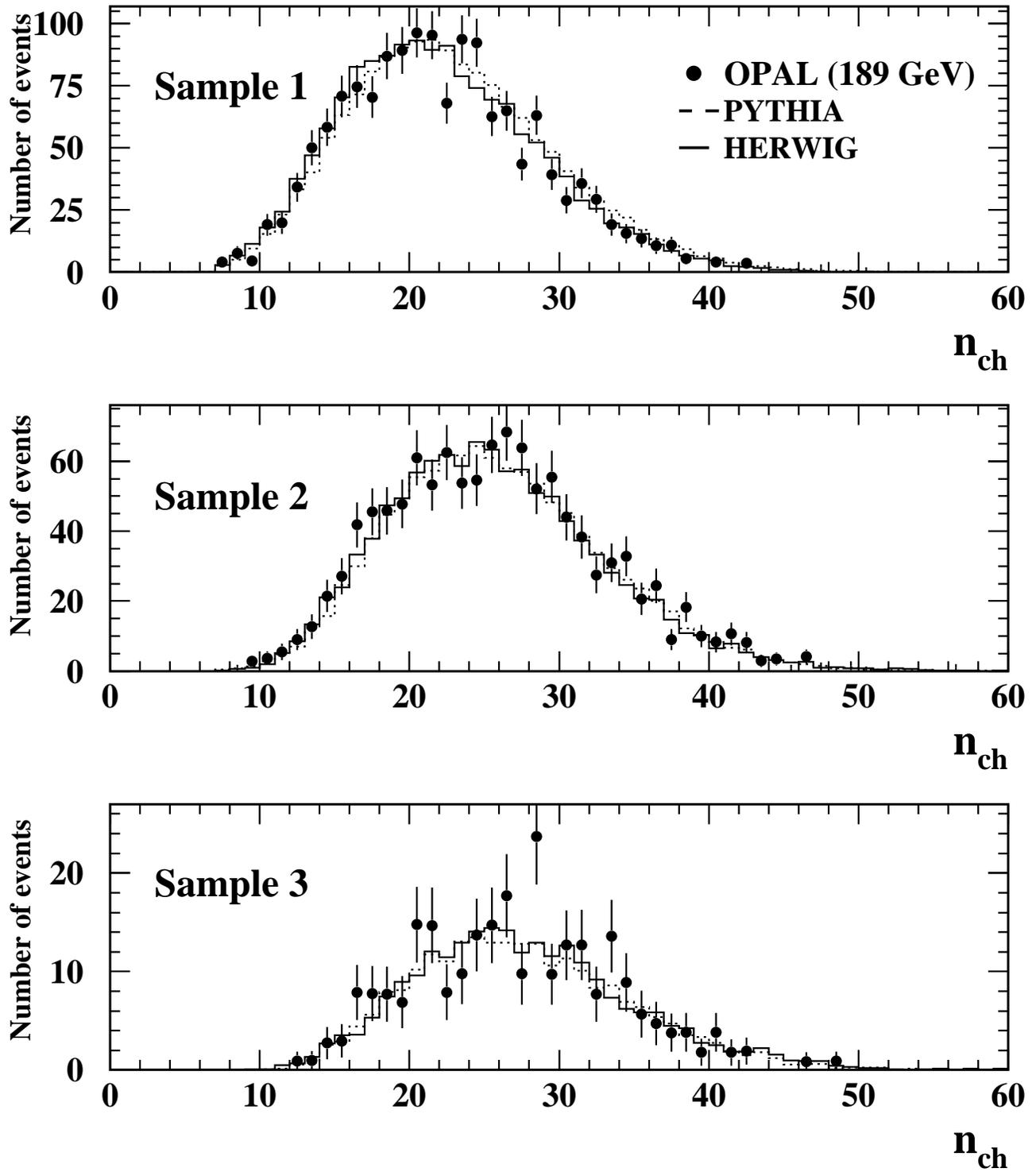}
\caption{\label{mult-distr-vs-mc}
Multiplicity distributions measured in the three samples 
at $\sqrt{s} = 189$~ GeV (solid points), after background subtraction, compared to  
the predictions from the PYTHIA (dotted histograms) and HERWIG 
(open histograms) models, obtained from fully simulated $\qq$ events.}
\end{figure}

A reliable correction requires a good simulation of the data. 
As an example, we show in Figure~\ref{mult-distr-vs-mc} the observed
charged particle multiplicity distributions, background subtracted,
as measured in the three samples at 189 GeV.
In the same figure we also show the corresponding distributions obtained
from fully simulated Monte Carlo events generated with PYTHIA and HERWIG, normalised to
the data integrated luminosity.
Qualitatively both models reproduce the measured distributions reasonably well at 
this energy and at all the other energies considered in this paper.

\begin{figure}[tbp]
\centering
\vspace*{-4mm}
\includegraphics[width=1.0\textwidth]{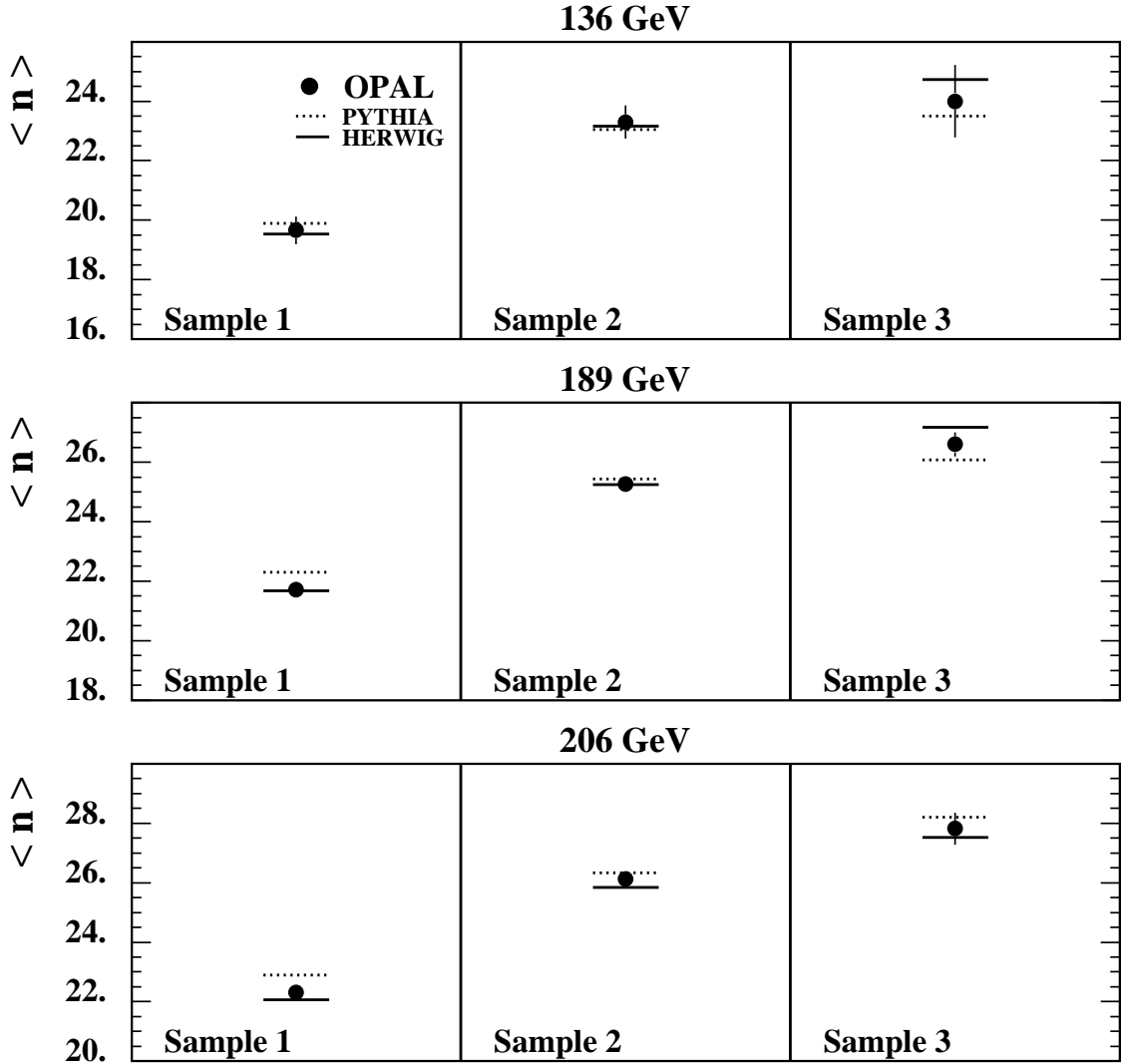}
\caption{\label{mean-values-vs-mc}
Mean charged particle multiplicities measured at  $\sqrt{s} = 136$, 189 and 206~GeV, after 
background subtraction (solid points), compared to the 
PYTHIA (dotted line) and HERWIG (solid line) model predictions.}
\end{figure}

The agreement between data and models 
is also shown in Figure~\ref{mean-values-vs-mc}, where the mean charged particle multiplicities 
as determined from the observed distributions after background subtraction are compared to
those predicted by PYTHIA and HERWIG at three different energies.
The errors are statistical only.
Again, within the statistical uncertainties, we observed a satisfactory agreement between data 
and predictions from both models at all centre-of-mass energies.
We therefore decided to correct our measured mean values using the coefficients $C_i$ computed, 
separately, with PYTHIA and HERWIG and quote the average of the two results as our reference value.
For PYTHIA the value of these coefficients varies from 1.19 to 1.22 
in Sample 1, from 1.05 to 1.10 in Sample 2 and from 1.08 to 1.10 in Sample 3,
depending on energy. 
The difference between these values and those predicted by HERWIG is typically less 
than 1\% and never exceeds 2\%.  

The system of equations (1) was then solved, at each energy, 
using the flavour fractions, $f_i$, predicted by the corresponding model.
The average between the two sets of results is presented in Table~\ref{table2} and defines 
our reference values.
The first uncertainty is statistical and the second is systematic.
The statistical uncertainties on $\dbl = \nb - \nl$ were computed taking into account the 
correlations between the measurements of $\nb$ and $\nl$.
The correlation coefficients are positive and vary from 0.30 to 0.44,
generally increasing with increasing energy.
As anticipated, the large uncertainties on the measured mean multiplicities of the $\cc$ events 
reflect our experimental inability to efficiently select c quark enriched samples. 
\section{Systematic uncertainties}

Several sources of possible systematic effects were considered.
\begin{itemize}

\item 
We have investigated a possible bias induced by the event selection. 
In particular an efficient suppression of the 4-fermion background 
was achieved by cutting on the $W_{\rm QCD}$ variable which, however,
also removes some $\qq$ events. 
We repeated the analysis using an alternative and independent cut.
By rejecting events with a thrust value $T<0.83$ we obtained the same 
level of background suppression.
Half of the difference between the reference and the varied results 
was taken as a systematic uncertainty.

\item
The subtraction of the residual 4-fermion background relies on cross sections 
and charged particle multiplicities as predicted by the grc4f/JETSET models.
By varying the predicted amount of background to be subtracted up and down by 5\%, 
slightly more than its measured uncertainty at $\sqrt{s} = 189$~GeV 
of 4\%~\cite{bib-sigma-bkg}, we have checked that the differences with respect to the 
reference results are negligible.
In previous OPAL analyses at 189 and 183 GeV~\cite{bib-mult-bkg}, it was demonstrated 
that the mean charged particle multiplicity of hadronic W decays measured in the data
and that predicted by the most commonly used hadronization models agreed within 1.1 
times the total experimental error, corresponding to $\pm 0.44$ on the multiplicity.
We varied the predicted mean charged particle multiplicity of the background by that 
amount, and take the difference between the reference and the varied results as a 
systematic error on our measurements. 
  
\item 
We have tested the stability of our results with respect to variations
of the cuts on the variable ${\cal L}$ which determine the high b quark purity 
in Sample~3 and the high uds quark purity in Sample~1.
The analysis was repeated, selecting Sample~3 with cuts at 0.7 and 0.9, or
selecting Sample~1 with cuts at 0.01 and 0.2, which lead to total variations of
about 10\% in the absolute b or uds quark purities of the samples.
For each case, the magnitude of the larger variation is taken to be the 
systematic trend.
In order to reduce statistical fluctuations in the estimate of the systematic
uncertainty, we take a statistically weighted average value as
a common systematic error at all energies.

\item
We have studied the systematic error associated with the simulation 
of the track resolution and its potential effect on our analysis through 
a change in performance of the b-tagging algorithms. 
A conservative estimate of this uncertainty was made by applying a smearing factor 
of 1.10 to the reconstructed $r-\phi$ and $z$ projections of the track's 
impact parameter and the polar and azimuthal angles, in the PYTHIA samples used 
for the standard analysis~\cite{bib-rb}.
The differences between the results with and without applying the smearing  
factor were taken as the systematic error. 

\item
Uncertainties associated with the modelling of b hadron production and decay in 
the simulation affect the predicted efficiencies and purities of the b-tagging procedure.
We have changed the $\epsilon_b$ parameter in the fragmentation function of Peterson et al.,
so as to vary the average scaled energy $<x_E>$ of b hadrons in the
range $<x_E> = 0.702 \pm 0.008$, as suggested by the LEP electroweak 
working group~\cite{bib-ewwg}.
The lifetimes of b mesons and baryons were varied by $\pm 0.02$~ps and $\pm 0.05$~ps,
respectively, based on the uncertainties on the measured values~\cite{bib-pdg}.
Finally, the average charged decay multiplicity of b hadrons was varied by $\pm 0.062$, 
reflecting the accuracy of the measurements by LEP experiments~\cite{bib-ewwg}.
The three effects considered were treated independently and for each of them the largest 
difference between the reference result and the varied result was taken as a systematic
uncertainty. 
At each energy the three uncertainties were added quadratically.
   
\item
To test the dependence of our results on the Monte Carlo model used to correct our data, 
we performed the analysis using, separately, both PYTHIA and HERWIG generators.
As already mentioned in the previous section, our reference results were taken as the 
average between the two sets of results, and we assign half of the difference as systematic 
uncertainty. 
\end{itemize}

The separate contributions to the systematic error on $\dbl$ are summarized in 
Table~\ref{table3}.
The total systematic uncertainty at a given energy was evaluated by adding all sources 
in quadrature.
In the last column of Table~\ref{table3} we show the systematic uncertainty of each source 
averaged over all energies.
The errors from each source were assumed to be completely correlated at different energies 
and were weighted by the corresponding statistical uncertainty.
The total averaged systematic error was again evaluated by adding all sources in quadrature.

\section{Comparison with QCD predictions and models}
%


According to perturbative QCD, soft gluon radiation from an 
energetic massive quark Q is suppressed 
inside a forward cone of half angle aperture $\Theta_0 = M_{\rm Q}/E_{\rm Q}$, the 
so-called Dead Cone~\cite{bib-dead-cone}.
Here $M_{\rm Q}$ is the heavy quark mass and $E_{\rm Q}$ its energy, and 
the relation holds if $E_{\rm Q} \gg M_{\rm Q} \gg \Lambda$, where $\Lambda$ 
represents the energy scale at which perturbation theory breaks down.
This phenomenon produces significant differences in the structure
of the soft gluon radiation emitted in light and heavy quark 
initiated jets.
Assuming the validity of the Local Parton Hadron Duality (LPHD) concept
a corresponding difference is also expected to be
present at the hadronic level. 
For $\epem$ annihilations a QCD calculation within the Modified 
Leading Log Approximation (MLLA)~\cite{bib-mlla}, assuming a b quark mass 
of 4.8 GeV/$c^2$, predicts a difference in mean charged particle 
multiplicity between $\bb$ and $\rm l\bar{l}$ events of $\dbl=5.5 \pm 0.8$, 
independent of energy~\cite{bib-schumm}. 
The quoted uncertainty is of experimental origin, while the uncertainty 
due to (energy-independent) missing higher order corrections 
is estimated to be about one unit.

More recently, the result of an improved calculation 
was published~\cite{bib-petrov} in terms of a conservative upper bound for 
$\dbl$, which was found to lie in the range 3.7 to 4.1 depending on the b quark 
mass, $m_{\rm b}$, assumed to be between 5.3 and 4.7 GeV, respectively.
In the same publication there was an attempt to evaluate more precisely 
the value of $\dbl$ at $\sqrt{s} = 91$ GeV, which gave $\dbl=3.68$ for 
$m_{\rm b} = 4.8$ GeV/$c^2$.
However, there is no general consensus~\cite{bib-khoze-ochs,bib-mult-lep2-th} 
on the theoretical consistency of the approach followed in~\cite{bib-petrov}, 
and it is still unclear whether a real improvement of the MLLA prediction  
has been achieved.
    
An independent upper limit of $\dbl<4$ was obtained from phenomenological
arguments and published in~\cite{bib-dedeus}. 


A more phenomenological approach, the \naive\ 
model~\cite{bib-rowson,bib-naive-model},  assumes that in heavy quark initiated 
events the multiplicity of light hadrons produced along with 
the heavy hadrons is the same as the total multiplicity of light quark 
initiated events produced at a centre-of-mass energy 
corresponding to the energy left behind after the heavy quarks have fragmented.
In this framework one expects the value of $\dbl$ to decrease with 
increasing energy.
There are several variations of this model which lead, however,
to only slightly different predictions.
We have used a form 
\begin{eqnarray}
\dbl=2\NQdec + \int\int N(\sqrt{(1-\xQ)(1-\xQb)}\sqrt{s})
f(\xQ)f(\xQb)\mathrm{d}\xQ
\mathrm{d}\xQb -N(\sqrt{s})
\end{eqnarray}
where $N(\sqrt{s})=2.554+0.1252\times\exp(2.317\sqrt{\ln \sqrt{s}})$ is a
parameterization of the world mean charged particle multiplicity
data, corrected to remove the effects of heavy quark production~\cite{bib-koetke},
$\xQ$ and $\xQb$ are the fractions of the beam energy carried by the 
heavy hadrons, $\NQdec$ is the decay mean multiplicity of the heavy hadrons,
and $\sqrt{s}$ is the centre-of-mass energy. 
We approximated the fragmentation function $f(\xQ)$ for b quarks by a normalized 
Peterson function with a mean of $0.70$, a conservative uncertainty of 
$\pm 0.02$ and assumed $2\NQdec=11.0 \pm 0.2$~\cite{bib-schumm}.

\begin{figure}[tbp]
\centering
\vspace*{-4mm}
\includegraphics[width=1.0\textwidth]{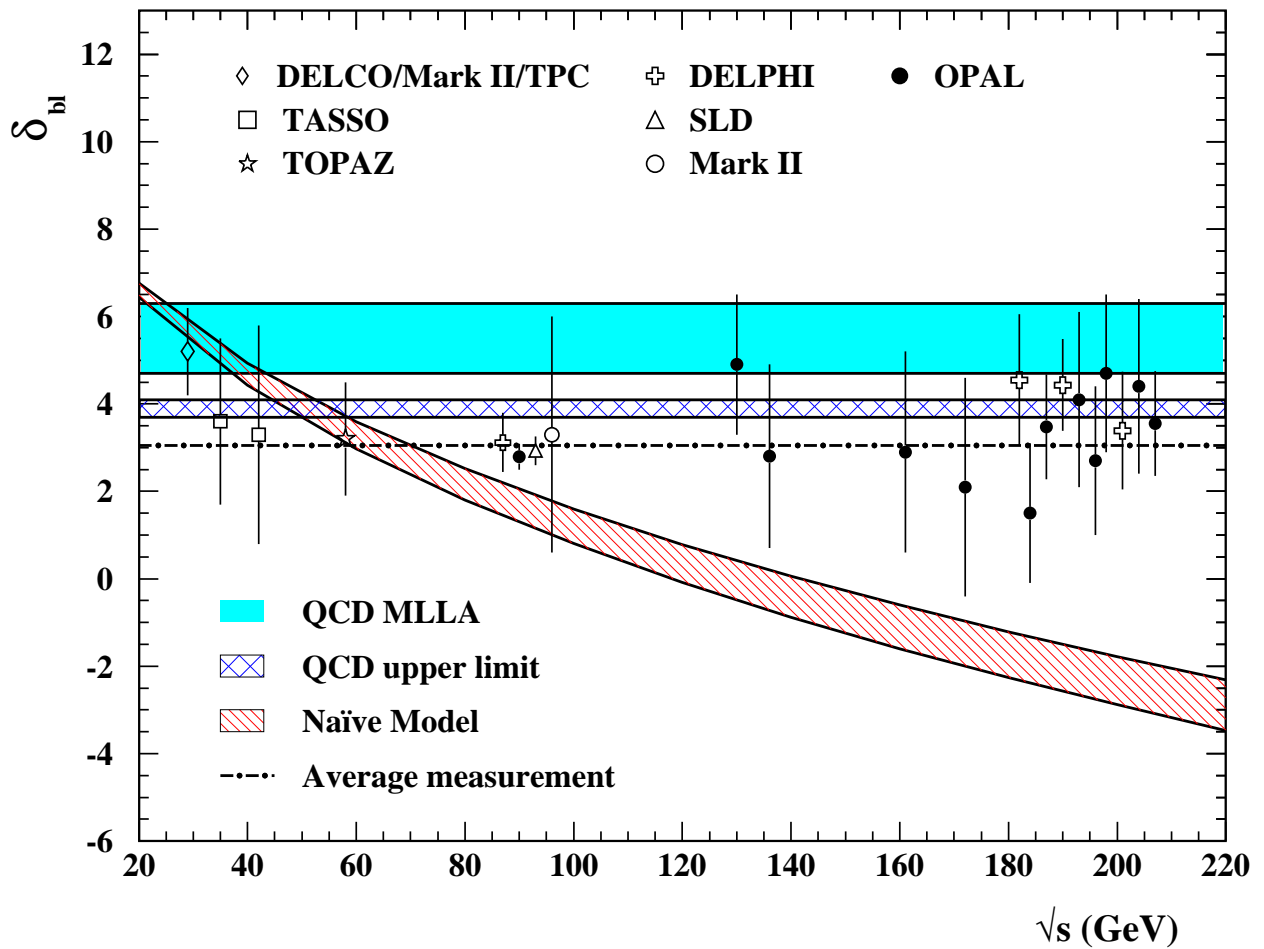}
\caption{\label{dbl_vs_energy}
The difference in mean charged particle multiplicity between $\bb$ 
and $\light$ events, $\dbl$, as a function of centre-of-mass energy. 
The data points show the experimental measurements and the total 
error, and those around  $\sqrt{s} = 91$, 183, 189 and 200~GeV have 
been separated horizontally for clarity.
The original MLLA prediction~\cite{bib-schumm} is shown as 
a shaded area to include the errors of experimental origin on this prediction,
not including missing higher order corrections.
The cross-hatched area corresponds to the QCD upper limits as 
calculated in~\cite{bib-petrov}.
The single hatched area represents the \naive\ model 
prediction~\cite{bib-rowson,bib-naive-model},
while the dash-dotted line is the combined result from all the
measurements, as discussed in section 6.}
\end{figure}

In Figure~\ref{dbl_vs_energy} we show our results on $\dbl$
as a function of energy, together with all 
previously published results~\cite{bib-rowson,bib-mark2,bib-opal-lep1,bib-sld,
bib-delphi-lep1,bib-low-e-dbl,bib-delphi-lep2}.  
Statistical and systematic uncertainties have been added in quadrature.

In the same figure, the theoretical predictions are also shown.
The striking difference between the QCD predictions (shaded 
area~\cite{bib-schumm} and cross-hatched area~\cite{bib-petrov}) and the 
prediction of the \naive\ model (single hatched area) is particularly 
evident at the highest LEP2 energies. 
One can see that the previously published results at the \zb\ peak energy 
and below did not allow a clear discrimination between the models.
Overall they are consistent with an energy independent behaviour,
but the \naive\ model could not be ruled out.
The results published by the DELPHI Collaboration at three different 
LEP2 energies~\cite{bib-delphi-lep2} showed a clear inconsistency
with the predictions of the \naive\ model. 

Our new results are consistent with those published by 
DELPHI~\cite{bib-delphi-lep2}, 
cover a much wider energy range and provide even stronger 
evidence of the inadequacy of this model.  

A linear fit to our eleven data points, considering only their statistical 
uncertainties, yields a slope of 
$0.000 \pm 0.018$ ($\chi^2/{\rm dof} = 0.59$), completely consistent 
with the QCD prediction of energy independence.
Repeating the fit to a constant value or, equivalently, combining our results
at a luminosity weighted average energy of 195 GeV, gives
$\dbl = 3.44 \pm 0.40$~(stat)~$\pm 0.89$~(syst). 
The overall systematic uncertainty of 0.89 was calculated assuming that 
each source of systematic uncertainty is fully correlated between energy points 
(see Table~\ref{table3}).
This value is consistent with the published OPAL result at 91 GeV~\cite{bib-opal-lep1} 
of $\dbl = 2.79 \pm 0.30$ (total error) and with the value of 
$2.99 \pm 0.20$ ($\chi^2/{\rm dof} = 0.79$) obtained from the corresponding fit to 
all published results up to and including 91 GeV, assuming that the measurements are completely 
uncorrelated. 
A weighted average including results from low energy data, 
LEP1 and LEP2 gives  $\dbl = 3.05 \pm 0.19$, which is shown in 
Figure~\ref{dbl_vs_energy} as a dash-dotted line. 
In this average we have assumed that the systematic errors of the DELPHI 
measurements at LEP2 are completely correlated between energy points.
 
One can also see from Figure~\ref{dbl_vs_energy} that the MLLA+LPHD 
prediction~\cite{bib-schumm} of $5.5 \pm 0.8$~(exp) is higher than  
the experimental results, even considering the additional theoretical uncertainty 
of about 1~unit due to missing higher order corrections. 
The upper bounds calculated in~\cite{bib-petrov} and in~\cite{bib-dedeus} are 
consistent with 
the measurements. 

\section{Conclusions}
We have measured the mean charged particle multiplicities for $\bb$, $\cc$ 
and $\light$ events at all energies collected by OPAL above the \zb\ peak,
and in particular we have determined the differences between the mean
multiplicity of b and uds initiated events, $\dbl=\nb-\nl$.

Our results are presented in Table~\ref{table2}
and are in agreement with previously published results~\cite{bib-delphi-lep2} at 
three LEP2 energies. 
Our data alone, which fairly uniformly cover a wide energy range, strongly support 
the QCD prediction of the energy independence of $\dbl$, leading to a combined result  
of 
\[\dbl = 3.44 \pm 0.40 ({\rm stat}) \pm 0.89 ({\rm syst)}\]
at a luminosity weighted average centre-of-mass energy of 195 GeV.
The consistency of the experimental results over the entire range 
from $\sqrt{s} = 29$ to $\sqrt{s} = 206$~GeV strengthens this conclusion even further.   

The \naive\ model, which assumes that the multiplicity accompanying 
the decay of a heavy quark is independent of the mass of the quark itself,
is strongly disfavoured.
\section*{Acknowledgements}
We particularly wish to thank the SL Division for the efficient operation
of the LEP accelerator at all energies and for their close cooperation with
our experimental group.  In addition to the support staff at our own
institutions we are pleased to acknowledge the  \\
Department of Energy, USA, \\
National Science Foundation, USA, \\
Particle Physics and Astronomy Research Council, UK, \\
Natural Sciences and Engineering Research Council, Canada, \\
Israel Science Foundation, administered by the Israel
Academy of Science and Humanities, \\
Benoziyo Center for High Energy Physics,\\
Japanese Ministry of Education, Culture, Sports, Science and
Technology (MEXT) and a grant under the MEXT International
Science Research Program,\\
Japanese Society for the Promotion of Science (JSPS),\\
German Israeli Bi-national Science Foundation (GIF), \\
Bundesministerium f\"ur Bildung und Forschung, Germany, \\
National Research Council of Canada, \\
Hungarian Foundation for Scientific Research, OTKA T-029328, 
and T-038240,\\
The NWO/NATO Fund for Scientific Reasearch, the Netherlands.\\
\clearpage
%
\renewcommand{\arraystretch}{1.10}
\begin{table}[htb]
\begin{center}
\begin{tabular}{|c||c|c|c|c|c|}
\hline
$\sqrt{s}$  & $\rm L_{int}$ & N. evts. & Sample 1            & Sample 2            & Sample 3       \\
(GeV)       & ($\rm pb^{-1}$) &   & $\rm N_{evt}$ (4-f bkg.) & $\rm N_{evt}$ (4-f bkg.) & $\rm N_{evt}$ (4-f bkg.) \\
            &     &    & b~~~~~c~~~~~uds             & b~~~~~c~~~~~uds             & b~~~~~c~~~~~uds             \\
\hline
\hline
 130  &    5.6 &  321 & 171 (negl.)     & 111 (negl.)     & 39 (negl.)     \\ 
      &        &      & 1.2~~18.3~~ 80.5   & 17.2~~31.0~~51.8  & 93.1~~5.5~~1.4   \\ \cline{2-6} 
 136  &    6.0 &  312 & 164 (negl.)     & 117 (negl.)     & 31 (negl.)     \\ 
      &        &      & 1.2~~19.1~~79.7   & 17.9~~30.9~~51.2  & 91.6~~5.8~~2.6   \\ \cline{2-6} 
 161  &   10.0 &  289 & 148 ($2.6\%$)   & 105 ($2.8\%$)   & 29 (negl.)     \\ 
      &        &      & 1.4~~21.0~~77.6   & 16.9~~32.3~~50.8  & 89.4~~7.1~~3.5   \\ \cline{2-6} 
 172  &   10.4 &  235 & 121 ($6.2\%$)   & 76 ($6.2\%$)    & 25 (negl.)     \\ 
      &        &      & 1.6~~21.3~~77.1   & 19.3~~32.6~~48.1  & 91.3~~6.5~~2.2   \\ \cline{2-6} 
 183  &   57.2 & 1016 & 463 ($11.6\%$)  & 367 ($8.3\%$)   & 90 ($2.2\%$)   \\ 
      &        &      & 1.7~~22.5~~75.8   & 19.2~~33.2~~47.6  & 91.1~~6.3~~2.6   \\ \cline{2-6} 
 189  &  181.8 & 3223 & 1493 ($12.0\%$) & 1153 ($8.6\%$)  & 259 ($2.6\%$)  \\ 
      &        &      & 1.7~~23.6~~74.7   & 18.7~~32.7~~48.6  & 90.3~~6.8~~2.9   \\ \cline{2-6} 
 192  &   26.9 &  492 & 235 ($11.6\%$)  & 165 ($8.8\%$)   & 44 ($2.2\%$)   \\ 
      &        &      & 1.7~~22.9~~75.4   & 19.4~~32.3~~48.3  & 89.7~~7.9~~2.4   \\ \cline{2-6} 
 196  &   54.8 & 1086 & 540 ($10.6\%$)  & 354 ($8.5\%$)   & 93 ($2.1\%$)   \\ 
      &        &      & 1.5~~23.1~~75.4   & 19.4~~33.0~~47.6  & 90.1~~6.7~~3.2   \\ \cline{2-6} 
 200  &   74.3 & 1137 & 511 ($14.3\%$)  & 419 ($8.9\%$)   & 79 ($2.5\%$)   \\ 
      &        &      & 1.8~~23.5~~74.7   & 19.5~~32.7~~47.8  & 89.4~~6.4~~4.2   \\ \cline{2-6} 
 202  &   37.1 &  538 & 238 ($14.1\%$)  & 193 ($8.1\%$)   & 50 ($2.0\%$)   \\ 
      &        &      & 1.8~~22.7~~75.5   & 18.8~~32.9~~48.3  & 90.0~~7.1~~2.9   \\ \cline{2-6} 
 206  &  218.0 & 2893 & 1319 ($15.0\%$) & 1036 ($8.7\%$)  & 201 ($2.9\%$)  \\ 
      &        &      & 2.2~~23.7~~74.1   & 20.1~~31.8~~48.1  & 89.4~~6.8~~3.8   \\ \cline{2-6} 
\hline
\end{tabular}
\end{center}
\caption{\label{table1}
The integrated luminosity, $\rm L_{int}$, collected at each energy and the total number 
of events after all selection cuts are shown in the first three columns. 
The number of events in each sample after the subtraction of the residual 4-fermion background, 
the fraction of events due to background which was subtracted (in parentheses) and the 
percentage of $\bb$, $\cc$ and $\light$ events as predicted by PYTHIA are also shown.}
\end{table}
\begin{table}[t]
\begin{center}
\begin{tabular}{|c||c|c|c|c|}  
\hline
$\sqrt{s}$ (GeV) &  \nb                      & \nc                        & \nl                       & \dbl            \\      
\hline 
\hline
 130 & 25.9  $\pm$ 1.2  $\pm$ 0.4 & 31.5 $\pm$ 4.5 $\pm$ 1.7 & 21.0  $\pm$ 1.4  $\pm$ 0.2  &  4.9 $\pm$ 1.5  $\pm$ 0.4 \\ \cline{2-5}
 136 & 25.7  $\pm$ 1.6  $\pm$ 0.5 & 26.2 $\pm$ 4.0 $\pm$ 2.1 & 23.0  $\pm$ 1.4  $\pm$ 0.8  &  2.8 $\pm$ 1.8  $\pm$ 1.0 \\ \cline{2-5}
 161 & 24.1  $\pm$ 1.6  $\pm$ 0.5  & 36.5 $\pm$ 5.0 $\pm$ 2.0 & 21.1  $\pm$ 1.9  $\pm$ 0.9  &  2.9 $\pm$ 2.0  $\pm$ 1.1 \\ \cline{2-5}
 172 & 28.8  $\pm$ 2.1  $\pm$ 0.5 & 21.7 $\pm$ 5.5 $\pm$ 1.9 & 26.8  $\pm$ 2.0  $\pm$ 0.8 &  2.1 $\pm$ 2.3  $\pm$ 1.0\\ \cline{2-5}
 183 & 28.3  $\pm$ 1.1  $\pm$ 0.5 & 25.8 $\pm$ 2.8 $\pm$ 2.0 & 26.8  $\pm$ 1.2  $\pm$ 1.0 &  1.5 $\pm$ 1.3  $\pm$ 1.0\\ \cline{2-5}
 189 & 28.89 $\pm$ 0.60 $\pm$ 0.49 & 29.8 $\pm$ 1.8 $\pm$ 1.9 & 25.41 $\pm$ 0.72 $\pm$ 0.76 & 3.48 $\pm$ 0.77 $\pm$ 0.98\\ \cline{2-5}
 192 & 28.5  $\pm$ 1.2  $\pm$ 0.7 & 33.1 $\pm$ 4.4 $\pm$ 2.1 & 24.4  $\pm$ 1.7  $\pm$ 0.9  &  4.1 $\pm$ 1.7  $\pm$ 1.1 \\ \cline{2-5}
 196 & 31.3  $\pm$ 1.4  $\pm$ 0.6 & 23.6 $\pm$ 3.3 $\pm$ 1.9 & 28.6  $\pm$ 1.3  $\pm$ 0.9 &  2.7 $\pm$ 1.5  $\pm$ 0.9 \\ \cline{2-5}
 200 & 30.3  $\pm$ 1.1  $\pm$ 0.6 & 31.0 $\pm$ 3.2 $\pm$ 2.9 & 25.6  $\pm$ 1.3  $\pm$ 0.8  &  4.7 $\pm$ 1.3  $\pm$ 1.3 \\ \cline{2-5}
 202 & 29.9  $\pm$ 1.6  $\pm$ 0.6 & 34.2 $\pm$ 4.7 $\pm$ 2.3 & 25.5  $\pm$ 1.9  $\pm$ 0.6  &  4.4 $\pm$ 1.9  $\pm$ 0.5 \\ \cline{2-5}
 206 & 30.08 $\pm$ 0.85 $\pm$ 0.63 & 30.4 $\pm$ 2.2 $\pm$ 2.0 & 26.53 $\pm$ 0.83 $\pm$ 1.10 & 3.55 $\pm$ 0.92 $\pm$ 0.84\\ \cline{2-5}
\hline
\end{tabular}
\end{center}
\caption{\label{table2}
Corrected mean charged particle multiplicities for $\bb$, $\cc$, $\rm l\bar{l}$ 
($\rm l\bar{l} = \light$) events and the difference $\dbl = \nb - \nl$ , measured 
at different energies.
The first error is  statistical and the second systematic in each case.}
\end{table}
\begin{table}[t]
\begin{center}
\begin{tabular}{|l||l|l|l|l|l|l|l|l|l|l|l||l|}  
\hline
$\sqrt{s}$ (GeV)  & 130  & 136  & 161  & 172  & 183  & 189  & 192  & 196  & 200  & 202  & 206 & Mean \\
\hline 
\hline
4-f rejection     & ~~--   & ~~--   & 0.40 & 0.12 & 0.35 & 0.02 & 0.89 & 0.50 & 0.14 & 0.13 & 0.18 & ~0.22 \\
4-f subtraction   & ~~--   & ~~--   & 0.07 & 0.05 & 0.04 & 0.09 & 0.10 & 0.07 & 0.12 & 0.11 & 0.09 & ~0.09 \\
b purity          & 0.23   & 0.23   & 0.23 & 0.23 & 0.23 & 0.23 & 0.23 & 0.23 & 0.23 & 0.23 & 0.23 & ~0.23 \\
uds purity        & 0.17   & 0.17   & 0.17 & 0.17 & 0.17 & 0.17 & 0.17 & 0.17 & 0.17 & 0.17 & 0.17 & ~0.17 \\
Track resolution  & 0.09   & 0.26   & 0.21 & 0.44 & 0.69 & 0.55 & 0.46 & 0.44 & 0.94 & 0.11 & 0.48 & ~0.49 \\
b modelling       & 0.29   & 0.29   & 0.27 & 0.26 & 0.26 & 0.26 & 0.25 & 0.25 & 0.24 & 0.23 & 0.21 & ~0.25 \\
Model dependence  & 0.11   & 0.90   & 0.89 & 0.76 & 0.54 & 0.71 & 0.40 & 0.47 & 0.86 & 0.29 & 0.55 & ~0.60 \\
\hline
TOTAL             & 0.43   & 1.02   & 1.07 & 0.97 & 1.02 & 0.98 & 1.15 & 0.90 & 1.34 & 0.51 & 0.84 & ~0.89 \\ 
\hline
\end{tabular}
\end{center}
\caption{\label{table3}
Contributions from different sources to the systematic uncertainty on the $\dbl$ measurements.
In the last column we show the average of the systematic uncertainties over all energies.
The errors from each source were assumed to be fully correlated at different energies, and 
were weighted by the corresponding statistical uncertainty. The total averaged uncertainty 
was obtained by adding all the contributions in quadrature.}
\end{table}
%
%
%
\clearpage
\end{document}